\documentstyle[11pt,psfig,twoside]{article}

\begin{document}

\newcommand{\dd}{\,{\rm d}}
\newcommand{\ie}{{\it i.e.},\,}
\newcommand{\etal}{{\it et al.\ }}
\newcommand{\eg}{{\it e.g.},\,}
\newcommand{\cf}{{\it cf.\ }}
\newcommand{\vs}{{\it vs.\ }}
\newcommand{\zdot}{\makebox[0pt][l]{.}}
\newcommand{\up}[1]{\ifmmode^{\rm #1}\else$^{\rm #1}$\fi}
\newcommand{\dn}[1]{\ifmmode_{\rm #1}\else$_{\rm #1}$\fi}
\newcommand{\upd}{\up{d}}
\newcommand{\uph}{\up{h}}
\newcommand{\upm}{\up{m}}
\newcommand{\ups}{\up{s}}
\newcommand{\arcd}{\ifmmode^{\circ}\else$^{\circ}$\fi}
\newcommand{\arcm}{\ifmmode{'}\else$'$\fi}
\newcommand{\arcs}{\ifmmode{''}\else$''$\fi}
\newcommand{\MS}{{\rm M}\ifmmode_{\odot}\else$_{\odot}$\fi}
\newcommand{\RS}{{\rm R}\ifmmode_{\odot}\else$_{\odot}$\fi}
\newcommand{\LS}{{\rm L}\ifmmode_{\odot}\else$_{\odot}$\fi}

\newcommand{\Abstract}[2]{{\footnotesize\begin{center}ABSTRACT\end{center}
\vspace{1mm}\par#1\par
\noindent
{~}{\it #2}}}

\newcommand{\TabCap}[2]{\begin{center}\parbox[t]{#1}{\begin{center}
  \small {\spaceskip 2pt plus 1pt minus 1pt T a b l e}
  \refstepcounter{table}\thetable \\[2mm]
  \footnotesize #2 \end{center}}\end{center}}

\newcommand{\TableSep}[2]{\begin{table}[p]\vspace{#1}
\TabCap{#2}\end{table}}

\newcommand{\TableFont}{\footnotesize}
\newcommand{\TableFontIt}{\ttit}
\newcommand{\SetTableFont}[1]{\renewcommand{\TableFont}{#1}}

\newcommand{\MakeTable}[4]{\begin{table}[htb]\TabCap{#2}{#3}
  \begin{center} \TableFont \begin{tabular}{#1} #4 
  \end{tabular}\end{center}\end{table}}

\newcommand{\MakeTableSep}[4]{\begin{table}[p]\TabCap{#2}{#3}
  \begin{center} \TableFont \begin{tabular}{#1} #4 
  \end{tabular}\end{center}\end{table}}

\newenvironment{references}%
{
\footnotesize \frenchspacing
\renewcommand{\thesection}{}
\renewcommand{\in}{{\rm in }}
\renewcommand{\AA}{Astron.\ Astrophys.}
\newcommand{\AAS}{Astron.~Astrophys.~Suppl.~Ser.}
\newcommand{\ApJ}{Astrophys.\ J.}
\newcommand{\ApJS}{Astrophys.\ J.~Suppl.~Ser.}
\newcommand{\ApJL}{Astrophys.\ J.~Letters}
\newcommand{\AJ}{Astron.\ J.}
\newcommand{\IBVS}{IBVS}
\newcommand{\PASP}{P.A.S.P.}
\newcommand{\Acta}{Acta Astron.}
\newcommand{\MNRAS}{MNRAS}
\renewcommand{\and}{{\rm and }}
\section{{\rm REFERENCES}}
\sloppy \hyphenpenalty10000
\begin{list}{}{\leftmargin1cm\listparindent-1cm
\itemindent\listparindent\parsep0pt\itemsep0pt}}%
{\end{list}\vspace{2mm}}

\def\TYLDA{~}
\newlength{\DW}
\settowidth{\DW}{0}
\newcommand{\dw}{\hspace{\DW}}

\newcommand{\refitem}[5]{\item[]{#1} #2%
\def\REFARG{#3}\ifx\REFARG\TYLDA\else, {\it#3}\fi
\def\REFARG{#4}\ifx\REFARG\TYLDA\else, {\bf#4}\fi
\def\REFARG{#5}\ifx\REFARG\TYLDA\else, {#5}\fi.}

\newcommand{\Section}[1]{\section{#1}}
\newcommand{\Subsection}[1]{\subsection{#1}}
\newcommand{\Acknow}[1]{\par\vspace{5mm}{\bf Acknowledgements.} #1}
\pagestyle{myheadings}

\def\thefootnote{\fnsymbol{footnote}}
\begin{center}
{\Large\bf The Optical Gravitational Lensing Experiment.\\
\vskip3pt
Short Distance Scale to the LMC.
\footnote{Based on  observations obtained with the 1.3~m Warsaw
telescope at the Las Campanas  Observatory of the Carnegie Institution
of Washington.}}
\vskip1cm
{\bf A.~~U~d~a~l~s~k~i$^1$,~~G.~~P~i~e~t~r~z~y~\'n~s~k~i$^1$,~~P.~~W~o~\'z~n~i~a~k$^2$,\\
~~M.~~S~z~y~m~a~{\'n}~s~k~i$^1$,~~M.~~K~u~b~i~a~k$^1$,
~~ and~~K.~~\.Z~e~b~r~u~\'n$^1$}
\vskip3mm
{$^1$Warsaw University Observatory, Al.~Ujazdowskie~4, 00-478~Warszawa, Poland\\
e-mail: (udalski,pietrzyn,msz,mk,zebrun)@sirius.astrouw.edu.pl\\
$^2$ Princeton University Observatory, Princeton, NJ 08544-1001, USA\\
e-mail: wozniak@astro.princeton.edu}
\vskip1cm
\end{center}

\Abstract{ We present {\it UBVI} photometry of the eclipsing binary
HV2274 -- the system which has been recently used for distance
determination to the LMC by Guinan \etal (1998). We determine the
interstellar reddening to the star, $E(B-V)=0.149\pm0.015$~mag,  based
on observed color indices of the star. This value is in excellent
agreement with the mean reddening towards HV2274 obtained from
photometry of the red clump stars in the surrounding field. The
reddening is almost twice as large as determined by Guinan \etal (1998).

We discuss the consequences of reddening underestimate. Most likely
HV2274 is located much closer with the distance modulus to the star and
the LMC: $m-M=18.22\pm0.13$~mag supporting the short distance scale to
the LMC. Such a distance modulus is in excellent agreement with the
recent distance determinations with RR~Lyr and red clump stars.}

\Section{Introduction}

Determination of the distance to the Large Magellanic Cloud is one of
the most important problems of the modern astrophysics. The
extragalactic distance scale based on Cepheids is tied to the LMC
distance (Kennicutt, Freedman \& Mould 1995) and therefore any error in
the distance determination to the LMC propagates  to extragalactic
distances.

Unfortunately the distance to the LMC has been a subject of controversy
for many years. The widely accepted distance modulus to this galaxy is
equal to $m-M=18.50\pm0.10$~mag based on Cepheid period-luminosity
relation (Laney \& Stobie 1994) and some determinations from the SN1987A
light echo (\eg Panagia \etal 1997). The latter method is in principle a
precise geometric technique. It suffers, however, in this case from
insufficient quality observations at crucial moments after supernova
eruption and many modeling assumptions.

On the other hand smaller, short distance modulus to the LMC was
determined from another standard candle -- RR~Lyr stars. The difference
amounts to about $0.2\div 0.4$~mag (Popowski \& Gould 1998, Udalski
1998a). Additional support for the short distance scale to the LMC was
provided by determination based on the red clump stars method proposed
by Paczy{\'n}ski \& Stanek (1998). Preliminary distance modulus obtained
with this method was found to be about 18.1 mag (Udalski \etal 1998,
Stanek, Zaritsky and Harris 1998)  with the final value of
$m-M=18.18\pm0.06$~mag (Udalski 1998b). It should be noted that the red
clump stars are the only standard candle calibrated with precise
parallaxes of hundreds of nearby stars measured by Hipparcos.

The possible solution of the LMC distance controversy should be found by
distance determination with another reliable standard candle. Eclipsing
binary stars seem to be the most promising candidates (Paczy{\'n}ski
1997). This method is largely geometric, free from possible population
effects influencing other standard candles.  Eclipsing binaries are very
common in the LMC, in particular large samples of such stars are found
as a by-product of microlensing searches (Grison \etal 1994, Alcock
\etal 1997). Accuracy of distance of a few percent should be achieved
when good quality observations are available.

Recently Guinan \etal (1998) presented the first determination of the
distance to the LMC with eclipsing binary HV2274. This system seems to
be well suited for distance determination. Using photometry of Watson
\etal (1992), radial velocity curve determined from HST observations and
temperatures of both components as well as interstellar reddening
derived from the blue and UV spectrum from the HST/FOS, Guinan \etal
(1998) obtained the distance modulus of HV2274 equal to
$m-M=18.47\pm0.07$~mag consistent with the long distance to the LMC.

A closer look at the parameters of HV2274 determined by Guinan \etal
(1998), in particular the interstellar reddening, raises some doubts on
robustness of method they applied. $E(B-V)=0.083\pm0.006$~mag  obtained
by Guinan \etal (1998) is very low and with the average foreground
(Galactic) reddening towards the LMC: $E(B-V)_{\rm for}=0.075$~mag
(Schlegel, Finkbeiner \& Davis 1998) it would suggest that either the
star is located in the very front of the LMC (contrary to Guinan \etal
claim that HV2274 is in the disk) or that there is a low extinction
window in this line-of-sight. On the other hand HV2274 is located only
$19'$ south from the edge of the reddening map of Harris, Zaritsky \&
Thompson (1997) which  suggests that in the region closest to HV2274 the
reddening is much higher: $E(B-V)\approx 0.15$~mag.

Because underestimated reddening leads to overestimated distance and the
Guinan \etal (1998) blue and UV observations are particularly sensitive
to reddening errors, we undertook a sub-project of the OGLE microlensing
search aiming at precise determination of the reddening to HV2274. This
paper presents the results leading to conclusion that HV2274 is most
likely located at the short LMC distance.

\Section{Observations}

{\it UBVI} photometry of HV2274 was  carried out as a sub-project of the
second phase of the OGLE microlensing survey with the 1.3-m Warsaw
telescope at the Las Campanas Observatory which is operated by the
Carnegie Institution of Washington. Single chip "first generation" CCD
camera with $2048\times 2049$ pixel SITe thin detector in "still frame"
mode was used. More details about the instrumental setup can be found in
Udalski, Kubiak \& Szyma{\'n}ski (1997).

HV2274 was observed on 11 photometric nights from Sep.~9, 1998 through
Sep.~25, 1998. On 7 nights only {\it BVI} photometry was obtained while
on the following 4 nights full {\it UBVI} set was collected. On each
night more than 20 standard stars from different Landolt (1992) fields
were also observed.

All frames were de-biased and flat-fielded with the standard OGLE data
pipeline. To achieve maximum precision we decided to make the aperture
photometry of HV2274. The star is well separated from the faint field
stars at the typical seeing of observations of 1.3 arcsec, and therefore
possible contamination of the aperture photometry is marginal for
this star. To derive {\it VI} photometry of the field stars we applied
identical procedure as described in Udalski (1998b). Calibrated {\it VI}
photometry of the field stars was compared with photometry of
LMC$\_$SC14N field (Udalski \etal 1998) which overlaps by about 70
pixels with HV2274 field. The mean differences of {\it V} and {\it
I}-band magnitudes were found to be about 0.01 mag ensuring that our
calibration procedures were performed correctly.

Transformation to the standard system was based on observations of a few
fields from Landolt (1992). The systematic error of the {\it V}-band
transformation for individual night should be smaller than 0.02~mag,
while for $B-V$ and $V-I$ colors -- smaller than 0.015~mag. For $U-B$
color the systematic error might be larger and we conservatively assume
it to be equal to 0.04~mag. This is caused by somewhat different
spectral response of the OGLE instrumental ultraviolet filter because of
somewhat steeper cut of the short wavelength ($\lambda<3500$\AA) side of
the {\it U}-band by the telescope field corrector made from BK7 Schott
glass. Nevertheless, the standard stars transform well to the standard
system. The color transformation coefficient for $U-B$ is equal to 1.17
(1.00 means the standard system) which compares to 0.96 and 0.98 for
$B-V$ and $V-I$, respectively. Typical $O-C$ differences for standard
stars are below 0.05~mag in $U-B$ in the range $-1.2<U-B<0.5$.

Table~1 lists observations of HV2274 collected during this program.

\MakeTable{ccccc}{12.5cm}{Photometry of HV2274}
{
\hline
\noalign{\vskip2pt}
JD hel. & $V$ & $U-B$ & $B-V$ & $V-I$ \\
$- 2451000$& & & & \\
\noalign{\vskip2pt}
\hline
\noalign{\vskip2pt}
65.8912 & 14.788 & $-$ & $-0.124$ & $-0.124$ \\
68.8513 & 14.800 & $-$ & $-0.122$ & $-0.140$ \\
69.8476 & 14.202 & $-$ & $-0.125$ & $-0.125$ \\
71.8598 & 14.472 & $-$ & $-0.141$ & $-0.100$ \\
74.8784 & 14.318 & $-$ & $-0.132$ & $-0.121$ \\
75.8239 & 14.162 & $-$ & $-0.130$ & $-0.123$ \\
75.8401 & 14.162 & $-$ & $-0.123$ & $-0.117$ \\
76.8473 & 14.173 & $-$ & $-0.137$ & $-0.132$ \\
77.8489 & 14.197 & $-0.902$ & $-0.123$ & $-0.141$ \\
79.8402 & 14.293 & $-0.915$ & $-0.121$ & $-0.118$ \\
80.8455 & 14.209 & $-0.918$ & $-0.137$ & $-0.127$ \\
81.8365 & 14.170 & $-0.884$ & $-0.134$ & $-0.132$ \\
\noalign{\vskip2pt}
\hline
}

\Section{Reddening to HV2274}

The interstellar reddening to HV2274 can directly be derived from {\it
UBVI} photometry of the star. The mean observed colors of HV2274 are as
follows: $(B-V)=-0.129\pm0.007$, $(V-I)=-0.125\pm0.011$ and
$(U-B)=-0.905\pm0.013$ where errors are the standard  deviations of
all measurements. Because the transformation coefficient of our $U-B$
color is worse than $B-V$ and $V-I$ colors  we assume conservatively
larger possible error of $\pm0.04$ for $U-B$ color.

The reddening can be determined from $(U-B)-(B-V)$ color-color diagram
with the most classical method -- by measuring the shift of the observed
position of the star along the reddening line from intrinsic, unreddened
colors of stars. Fig.~1 presents $(U-B)-(B-V)$ diagram for early
spectral type stars. The solid line represents position of unreddened
giants  and dash-dotted line position of main sequence stars. Both lines
are the linear fits to the Schmidt-Kaler (1982) calibrations. Because
the calibration of Schmidt-Kaler is based on Galactic objects which are
in general of higher metallicity than similar stars in the LMC we
calculated synthetic $U-B$ and $B-V$ colors for a grid of Kurucz (1992)
models of atmospheres for ${\rm log}~g=3.5$, temperatures $T_{\rm eff}$
ranging from 21000~K to 25000~K and metallicities [Fe/H]: 0.0~dex
(Galactic stars) and $-0.5$~dex (LMC stars). The differences between
$U-B$ and $B-V$ colors for two metallicities at the same temperature
were found to be 0.019~mag and 0.002~mag for $U-B$ and $B-V$ colors,
respectively. This result  confirms that the Galactic calibration can
safely be applied to LMC stars of early B spectral type.

The star in Fig.~1 indicates position of HV2274 and the arrow shows
direction of the mean reddening line for the LMC: $E(U-B)/E(B-V)=0.76$
(Fitzpatrick 1985). Dotted line marks the shift of HV2274 due to
reddening. The resulting reddening of HV2274 is equal to
$E(B-V)=0.149\pm0.015$~mag. We assumed here luminosity class III for
HV2274.

Black dots in Fig.~1 show positions of 11 OB-type stars  with colors
corresponding to the main sequence stars located within $3'$ from
HV2274. $E(B-V)$ reddening of these stars resulting from their observed
$U-B$ and $B-V$  color indices is equal to 0.128, 0.147, 0.170, 0.125,
0.127, 0.129, 0.137, 0.124, 0.182, 0.121 and 0.211~mag for stars 
$1-11$, respectively. Thus, the mean $E(B-V)$ reddening towards HV2274
is equal to 0.146~mag with the standard deviation of 0.028~mag.  We may
conclude that OB-type stars in the field around HV2274 are reddened
similarly to HV2274.

We also applied to our data the "Q-method" of reddening determination.
This method was used by Harris \etal (1997) to produce their reddening
map. The values of $Q_1$ and $Q_2$ parameters calculated for HV2274
according to Harris \etal (1997) prescription are equal to $-0.808$ and
$-0.829$, respectively. The $E(B-V)$ reddening can immediately be
derived from Fig.~2 of Harris \etal (1997). It is equal to 0.158~mag
from $Q_1$ parameter and 0.155~mag from $Q_2$. These values are somewhat
larger than the value from color-color diagram but one has to remember
that the zero reddening line of Harris \etal (1997) comes from other
source.  Both methods give very consistent results.

To verify if we do not make any large systematic error we performed one
more, entirely independent determination of reddening towards HV2274. It
was shown that the mean luminosity of the red clump stars in the {\it
I}-band is constant and independent of age for population of stars
2--10~Gyr old (Udalski 1998b). It only slightly depends on metallicity
(Udalski 1998a). Therefore the red clump stars can serve as a good
luminosity reference for reddening determination.  Indeed this method
was used by Stanek (1996) to determine map of extinction of the Baade's
Window in the Galactic bulge.

The top panel of Fig.~2 presents the $I-(V-I)$ color-magnitude diagram
for stars located in the area of $150\times 150$ arcsecs centered on
HV2274. In the lower panel the luminosity function of stars from the
range $19.2<I<17.2$ and $0.8<V-I<1.2$ is plotted with the fitted
function consisting of a Gaussian representing the  red clump stars and
parabola representing background stars of the red giant branch (Udalski
\etal 1998). The mean brightness of the red clump stars towards HV2274
is $\langle I \rangle =18.207$~mag with statistical uncertainty of
0.017~mag and the standard deviation of the Gaussian  $\sigma_{\rm
RC}=0.18$~mag. The mean luminosity of the red clump stars in the LMC is
equal to $\langle I_0\rangle =17.88$~mag at metallicity $-0.8$~dex
(Udalski 1998b) which after a small correction for metallicity according
to Udalski (1998a) calibration corresponds to $\langle I_0\rangle
=17.90$~mag at the mean metallicity of the field stars in the LMC equal
to $-0.6$~dex (Bica \etal 1998). Thus the mean extinction towards HV2274
in the {\it I}-band is equal to $A_I=0.307$~mag which corresponds to the
mean reddening in that direction of $E(B-V)=0.157$~mag. This value is in
very good agreement with our previous determination based on OB-type
stars.

The agreement between the mean reddening towards HV2274  and the value
from individual  determination to HV2274 is excellent and reassures that
our reddening determination was correct. Guinan \etal (1998) conclude
from similar radial velocities of the ISM lines and systemic velocity of
HV2274 that the star must be located in the LMC disk. Consistent
reddening of HV2274 and bulk of disk stars confirms that this is indeed
the case.

\Section{Discussion}

The value of interstellar reddening to HV2274,
$E(B-V)=0.149\pm0.015$~mag, determined in previous Section confirms our
guess expressed in the Introduction that Guinan \etal (1998)
underestimated the reddening while determining the distance to the LMC.
Their value of $E(B-V)=0.083\pm0.006$~mag is only half of that derived
from our {\it UBVI} photometry. This is not surprising because the
reddening was derived as one of many free parameters of model spectrum
fitted to HST data for HV2274. Any degeneracy between free parameters
contributes to very large uncertainties in final results.

Larger reddening than obtained by Guinan \etal (1998) has very important
consequences for their results because their spectra cover blue and UV
regions where~extinction is a sensitive function of reddening.
Unfortunately we do not have the HST spectrum of HV2274 at our disposal
and therefore we were unable to repeat the Guinan \etal (1998) procedure
with correct value of reddening. Nevertheless some conclusions are
evident.

Most importantly, the HST spectrum presented in Fig.~2 by Guinan \etal
(1998) must have been dereddened incorrectly. In the reddest part of
their spectrum ($\lambda\approx 4800$\AA) where extinction is the
smallest ($A_{4800}\approx 3.8E(B-V)$), the observed dereddened flux at
the Earth was underestimated by as much as 0.25~mag.

This excess of flux can only be removed in two ways. First, if the
effective temperature of both components of HV2274 was determined
incorrectly then the larger $T_{\rm eff}$ would make both components
brighter. Effective temperature would have to be higher by about 3000~K
to account for the excess as indicated by Kurucz atmosphere models.

On the other hand such large error in effective temperatures determined
by Guinan \etal (1998) does not seem to be likely. In the inset in their
Fig.~2 they present enlarged part of the spectrum around the Balmer jump
and Balmer lines with very good fit of the model spectrum. The depth of
the hydrogen lines and the Balmer jump relative to the continuum is a
sensitive function of gravity and effective temperature. Thus with the
gravity well constrained from the system parameters, ${\rm
log}~g\approx3.5$, good fit of the Balmer jump and depth of the Balmer
lines indicates good temperature determination. The effect of larger
extinction is not important here as the line and surrounding continuum
are affected in similar way.  For instance,  Kurucz models indicate that
increase of temperature from 23000~K to 26000~K  produces shallower
Balmer lines and Balmer jump by about 20\%. Therefore a good fit of the
model to the  observed spectrum by Guinan \etal (1998) suggests that
uncertainty of the effective temperature should not exceed  $\approx
1000$~K.  It should be also noted that the constant  colors at different
orbital phases,  confirm Guinan \etal (1998) result that temperatures of
both components of the system are almost identical.

The second possibility of removing excess of flux is obvious. Smaller
distance to HV2274 would make it brighter. If we assume that the system
parameters of HV2274 were determined correctly then the distance modulus
of the star derived by Guinan \etal (1998), $m-M=18.47$~mag,  must be
smaller by 0.25~mag, that is $m-M=18.22$~mag.

At shorter wavelengths -- in the UV part of the spectrum -- the excess
of flux must be higher than at $4800$\AA. We suppose that  this should
not pose any problem in fitting the HST spectrum of HV2274 with the
model. The UV extinction curve: its slopes, strength of UV bump etc,
were left by Guinan \etal (1998) as other free parameters when fitting
the spectrum. Thus less steep UV extinction curve can likely incorporate
the excess of flux in UV producing a fit not worse than presented by
Guinan \etal (1998).

Summarizing, the reddening to HV2274 was significantly underestimated by
Guinan \etal (1998) because of very likely degeneracy of UV extinction
curve and $E(B-V)$ reddening. The effective temperature of both
components of the system seems, however, to be determined correctly.
Therefore the most likely consequence of the underestimated reddening is
smaller distance modulus to HV2274 equal to $m-M=18.22\pm0.06$~mag with
the error resulting from reddening uncertainty. It is difficult to
estimate the error of the distance resulting from uncertainties of
stellar parameters and effective temperatures but we believe that it
should not exceed 0.1~mag.

The distance modulus of HV2274, $m-M=18.22\pm0.13$~mag, the star
located in the disk of the LMC, can be then assumed as the distance 
to the LMC. We do not apply here any geometric correction to HV2274
distance modulus  as Guinan \etal (1998) did because this does not seem
to be confirmed by observations of the red clump and RR~Lyr stars
located in fields on opposite sides of the LMC bar (Udalski \etal 1998,
Udalski 1998a).

It is worth noting here, that all recent  determinations of the
distance to the LMC with reliably calibrated methods and based on
massive photometry from microlensing surveys support the short distance
scale to the LMC. The distance to the LMC obtained from the mean {\it
I}-band brightness of the red clump stars corrected for population
effects is equal to $m-M=18.18\pm0.06$~mag (Udalski 1998b).

Another determination of the distance to the LMC was based on the mean
luminosity of the field RR Lyr stars in the LMC (Udalski 1998a). The
mean {\it V}-band brightness of the RR~Lyr stars was found to be
$\langle V_0\rangle = 18.86$~mag for 110 RR~Lyr stars located in fields
on opposite sides of the LMC bar. When this figure was derived precise
$E(B-V)$ reddening was not known for these fields and extrapolated
values were used. The reddening  was, however, slightly overestimated --
on average by 0.023~mag, as indicated by determination with the red
clump stars in similar way as in Section~3. Thus the mean {\it V}-band
brightness of the LMC field RR~Lyr stars is rather $\langle V_0\rangle =
18.93$~mag. With the most reliable calibration of the RR~Lyr absolute
luminosity based on statistical parallaxes of about 150 stars (Gould \&
Popowski 1998) the RR~Lyr distance modulus to the LMC is
$m-M=18.16\pm0.16$~mag.  All three independent methods: eclipsing binary
HV2274, red clump stars and RR~Lyr stars, give the short distance to the
LMC with very impressive agreement.

In the title of the first paper on determination of the LMC distance
with the red clump method (Udalski \etal 1998) we asked the question:
"Are the Magellanic Clouds 15\% closer than generally accepted?". The
answer seems to be more and more likely: Yes,  with all astrophysical
consequences: smaller distance to  extragalactic objects, larger Hubble
constant, fainter RR~Lyr stars and therefore older globular clusters.

{\bf Acknowledgements.} We would like to thank Prof.\ Bohdan Paczy\'nski
for many discussions. We thank Drs.\ K.Z.\ Stanek and D.\ Sasselov for 
comments and remarks on the paper. The  paper was partly supported by
the Polish KBN grant 2P03D00814 to A.\ Udalski.  Partial support for the
OGLE project was provided with the NSF grant AST-9530478 to
B.~Paczy\'nski.

\newpage

\centerline{\bf Figure Captions}

\vspace{1cm}

\noindent
Fig.~1. Location of HV2274 in $(U-B) - (B-V)$ color-color diagram
(star). Solid and dash-dotted lines represent position of unreddened
giants and main sequence stars, respectively. Arrow marks direction of
reddening. Black dots indicate positions of 11 OB-type stars from the
HV2274 field.

\noindent 
Fig.~2. Top panel: $I - (V-I)$ color magnitude diagram of stars in the
field around HV2274. Star indicates position of HV2274. Bottom panel:
Luminosity function of the red clump stars with fitted Gaussian
superimposed with parabola function. Bins are 0.07~mag wide.

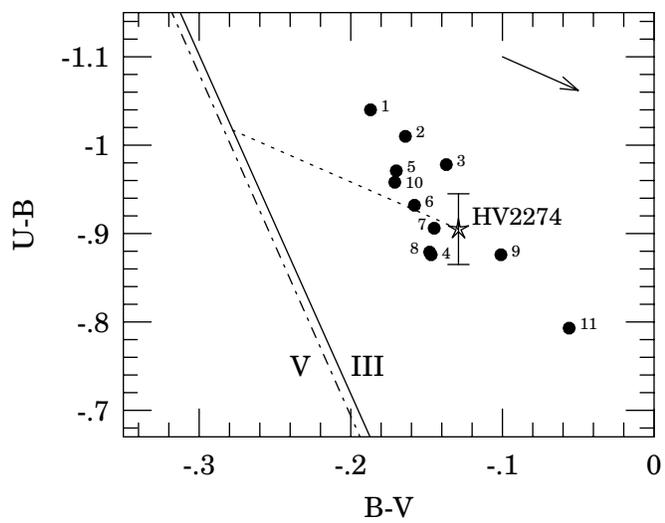
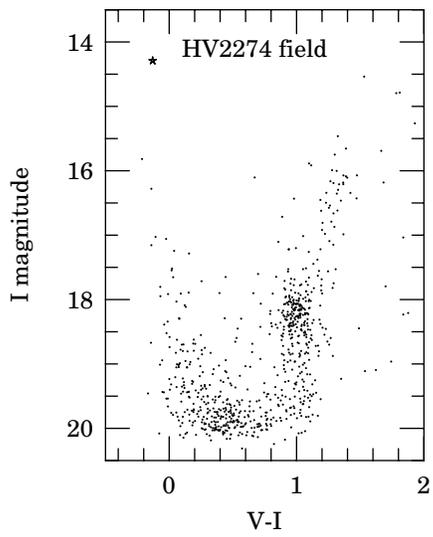
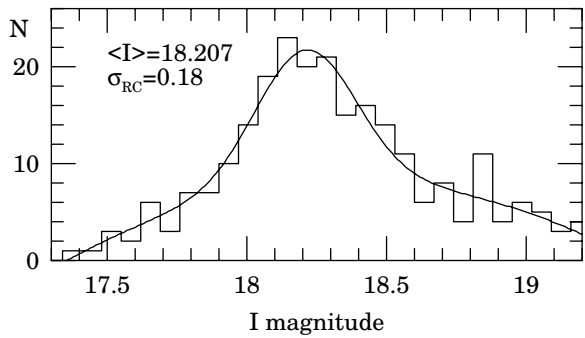
\end{document}